\documentclass[doublecol]{epl2}
\usepackage{amsmath, amssymb}
\usepackage{graphicx}

\def\bk{{\bf k}}

\newcommand{\be}{\begin{equation}}
\newcommand{\ee}{\end{equation}}
\newcommand{\beq}{\begin{eqnarray}}
\newcommand{\eeq}{\end{eqnarray}}

\title{Non-universal Casimir forces  at Bose-Einstein condensation of an ideal gas: effect of Dirichlet boundary conditions}
\shorttitle{Non-universal Casimir forces for Dirichlet boundary conditions} 

\author{M. Napi\'orkowski, \inst{1} J. Piasecki, \inst{1} \and J. W. Turner\inst{2}}
\shortauthor{M. Napi\'orkowski \etal}

\institute{                    
  \inst{1}  Institute of Theoretical Physics, Faculty of Physics, University of Warsaw, Pasteura 5, 02-093 Warszawa, Poland\\
  \inst{2} Universit\'e Libre de Bruxelles, Brussels, Belgium}

\abstract{We analyze the Casimir forces for an  ideal Bose gas enclosed  between two infinite parallel walls separated 
by the distance $D$. The walls are characterized by the Dirichlet boundary conditions. We show that if the thermodynamic state  with   Bose-Einstein condensate present is correctly approached along the path pertinent to the Dirichlet b.c. then the leading term describing the large-distance decay of thermal Casimir force between the walls is $\sim 1/D^{2}$ with a non-universal amplitude.  The next order correction is $\sim \ln D/D^3$. These observations remain in contrast with the decay law for both the periodic and Neumann boundary conditions for which the leading term is $\sim 1/D^3$ with a universal amplitude. We associate this discrepancy with the non-zero D-dependent positive value of the one-particle ground state energy  in the case of Dirichlet boundary conditions.}

\begin{document}

\maketitle 

\section{Introduction}
The properties of Casimir forces attracted broad attention in recent years. Various systems enclosed by walls of varied geometries have been studied both  experimentally and theoretically 
\cite{Krech1994,Most97,Kardar99,Bordag01,Brankov00,HHG2008,Gambassi09}. In particular, the critical region turned out to be of special interest because there the Casimir forces  are expected to gain, for large wall separations, a universal form. \\
A special role among systems investigated theoretically is played by the ideal Bose gas enclosed  between two infinite parallel walls. On  the one hand the bulk system displays the Bose-Einstein condensation while on the other hand various boundary conditions usually taken to be Dirichlet, periodic, or Neumann type  are straightforwardly incorporated  into  the analysis. The energy spectrum of the gas enclosed in a $L_{1} \times L_{2} \times L_{3}$ box  is composed of one-particle energy levels $\epsilon_{\bk} = \hslash^2 \bk^2/2m$ with the boundary-conditions dependent wave vector $\bk = (k_{1}, k_{2}, k_{3})$. In contradistinction to the periodic or Neumann boundary conditions where the ground state energy vanishes the Dirichlet boundary conditions imply the spectrum $k_{i} = \pi n_i/L_{i}$,  with $ n_i=1,2, ...$, $i=1,2,3$, so that $\bk^2\ge \pi^2(L_1^{-2}+L_2^{-2}+L_3^{-2})$ and the ground state energy $\epsilon_{G} > 0$. 
This important aspect of the energy spectrum has to be properly taken into account when discussing the Bose-Einstein condensation in the grand canonical ensemble parametrized by temperature $T$, chemical potential $\mu$, and volume $V=L_{1}L_{2}L_{3}$. 

The macroscopic occupation of the ground state takes place in  the thermodynamic limit only when $T < T_{c}$ and
 $\mu = 0 $. Here $T_{c}$ denotes the critical temperature defined by $\rho \lambda_{c}^3 = \zeta(3/2)$, where 
 $\rho = \lim\limits_{\infty} <N>/V$ is the number density of the gas evaluated in the thermodynamic limit, $\lambda_{c}$ denotes the thermal de Broglie wavelength $\lambda=h/\sqrt{2\pi m k_{B} T}$ evaluated at $T=T_{c}$, and $\zeta$ denotes the Riemann  zeta function. 
 The condensate density $\rho_{0}(T) = \lim\limits_{\infty}<n_G>/V$, where $<n_{G}>$ denotes the average number of particles in the ground state is non-zero for $T < T_{c}$ and $\mu =0$.  \\
 The {\it crucial} element of theoretical analysis of Bose-Einstein condensation consists in choosing the correct path along which the thermodynamic state of infinite system  with $\rho_0>0$  should be approached \cite{LL000,ZUK1977,LW1979}. Along this path taken at constant temperature 
$T < T_{c}$  the chemical potential cannot be kept constant because both  $V $ and 
$\mu$ must vary simultaneously in such a way that the equality  
\beq
 \label{defnG}
\frac{<n_{G}>}{V} =  \frac{1}{V}\, \left[\exp\left( \frac{\epsilon_{G}-\mu}{k_{B}T} \right)-1\right] ^{-1} = \rho_{0} . 
\eeq
holds when $V \rightarrow \infty $. 

The above defined procedure should be applied to the evaluation of both bulk and surface quantities. In particular, it should be followed when determining the Casimir forces for thermodynamic states characterized by the presence of the condensate. In the case of  Dirichlet boundary conditions,  where the ground state energy $\epsilon_{G} > 0$, the correctly chosen path  makes $\mu$ tend to zero when 
$V \rightarrow \infty$ according to the formula \cite{LL000,ZUK1977,LW1979}  
\beq
\label{mu001}
\mu=\mu^{D}(T,V)=\epsilon_{G} -  k_{B}T \ln\left(1+\frac{1}{V \rho_{0}}\right) .
\eeq
 If one chooses a different procedure in which one sets in eq.~(\ref{defnG}) $\mu=0$ first and only then takes the thermodynamic limit,  one is led to the erroneous conclusion about non-existence of Bose-Einstein condensation for Dirichlet boundary conditions \cite{R1976,LW1979}.  

\section{Casimir forces}
In this Section we consider the ideal Bose gas enclosed in a $L \times L \times D$ box with walls represented by Dirichlet boundary conditions. In this case the path defined in eq.~(\ref{mu001})  has the following form
\beq
\label{mu002}
\mu^{D}(T,L,D) = \nonumber \\ k_B T \left\{ \frac{\pi}{4} \left[2\left(\frac{\lambda}{L}\right)^2 + 
 \left(\frac{\lambda}{D}\right)^2\right] -  \ln\left(1+\frac{1}{L^2D \rho_{0}}\right) \right\}.
\eeq
We study the grand canonical free energy $\Omega(T,\mu,L,D)$ at $\mu = \mu^{D}(T,L,D)$ with the purpose of calculating  Casimir forces in the presence of condensate. The relevant quantity is then the surface contribution  obtained by subtracting from $\Omega(T,\mu,L,D)$ the bulk term $\Omega_{b}(T,L,D) = - L^2D\, p(T,0)$, where $p(T,0)$ denotes the Bose gas pressure at $\mu = 0$.

We focus here on the pair of walls each of area $L^2$ separated by distance $D$ and evaluate the surface free energy density
\beq
\label{fi01}
\Phi(T,D) = \lim\limits_{L \rightarrow \infty} \frac{\Omega(T,\mu^{D}(T,L,D),L,D)}{L^2} 
\eeq
describing the gas contained between two infinite walls.
The Casimir force ${F}_{C}(T,D)$ acting  between the  walls measures the rate of change with distance $D$ of the pure surface contribution 
\beq
\label{surf} 
\Phi_{s}(T,D)= \lim\limits_{L \rightarrow \infty}\frac{\Omega(T,\mu^D(T,L,D),L,D) - \Omega_b(T,L,D)}{L^2}=  \nonumber 
\eeq
\beq
\Phi(T,D)+Dp(T,0).
\eeq
It reads 
\beq
\label{Cas001}
{F}_{C}(T,D) =  -\frac{\partial \Phi_{s}(T,D)}{\partial D} = - \frac{\partial \Phi(T,D) }{\partial D}  +  p(T,0) \,.
\eeq

Our purpose is to evaluate $\Phi(T,D)$ and the corresponding Casimir force  
${F}_{C}(T,D)$  along the path defined in eq.~(\ref{mu002}),  and to determine its $D$ dependence for large $D$, i.e., $D \gg \lambda$. We then compare the large $D$ behavior of  $\Phi_{s}(T,D) $ with that of  the surface free energy density
\beq
\label{omegas02}
\Phi_{s}^0(T,D) =
 \lim\limits_{L \rightarrow \infty} \frac{\Omega(T,\mu = 0,L,D) - \Omega_{b}(T,L,D)}{L^2} .
\eeq
evaluated for fixed values of parameters $(T, \mu = 0,D)$. The latter procedure would thus correspond to setting
 $\mu=0$ first, and only then analyzing the large $D$ decay. As stressed before, it is known to give qualitatively wrong results in the bulk case for Dirichlet boundary conditions.   

 In order  to settle the above issue we evaluate $\Phi(T,D)$ 
using the energy spectrum under Dirichlet boundary conditions and equation (\ref{mu002}) (in the limit $L\to\infty$). One then arrives in a straightforward way at the formula
\beq
\label{fi02}
\Phi(T,D) = \nonumber \\ \frac{k_BT}{\lambda^2} \int\limits_{0}^{\infty} dw \sum\limits_{n=1}^{\infty} \ln\left\{1-\exp[-w- s^2(n^2-1)] \right\} \,,  
\eeq 
where $s = \sqrt{\pi}\lambda/2D$. 

In order to determine the leading  contributions to $\Phi(T,D)$ for $D \gg \lambda$  ($s \ll1$) 
we performed an asymptotic analysis which was not straightforward, rather cumbersome. So, for reasons of clarity we present here the final result first, and only then sketch the essential steps of our calculations.  

Our final result written in dimensionless form is as follows
\beq
\label{konc01}
\frac{\lambda^2}{k_BT}\Phi(T,D) = 
- \frac{2}{\sqrt{\pi}} \frac{1}{s} \zeta(5/2) + \frac{1}{2} \zeta(2)  \nonumber 
\eeq
\beq
-\frac{\sqrt{\pi}}{2}\, s \,\zeta(3/2) - 
s^2 \ln s -s^2 \left(\frac{\zeta(3)}{2\pi^2} -\frac{1}{2}\right) + o(s^2).
\eeq 
The first term on the rhs of eq.~(\ref{konc01}) is linear in $D $ and represents the bulk grand canonical free energy evaluated at $\mu=0$. It implies the well known formula  $p(T,0) = k_{B}T \zeta(5/2)/\lambda^3 $ \cite{LL000,ZUK1977}. \\
The remaining terms represent contributions to the surface free energy density  $\Phi_{s}(T,D) $ defined in eq.~(\ref{surf}). 
\beq
\label{omtd01}
\Phi_s(T,D)=2\sigma_{wg}(T)-
\frac{\pi k_{B}T}{4\lambda^2} \left[\zeta\left(\frac{3}{2}\right) \frac{\lambda}{D}+\left(\frac{\lambda}{D}\right)^2 \ln\left(\frac{\lambda}{D}\right)\right.  \nonumber 
\eeq
\beq
+ \left.\left(\frac{\zeta(3)}{2\pi^2} -\frac{1}{2} +\frac{1}{2} \ln\left(\frac{\pi}{4}\right)\right)\,\left(\frac{\lambda}{D}\right)^2 \right]+ 
o\left(\left(\frac{\lambda}{D}\right)^2\right) .
\eeq
The $D$-independent term $2\,\sigma_{wg}(T)$, where  $\sigma_{wg}(T) = \frac{k_{B}T}{4 \lambda^2} \zeta(2)$ is the coefficient of wall-gas surface tension evaluated at $\mu=0$ \cite{ZUK1977,NP2014}. \\

Our result should be compared with the asymptotic behavior of the surface free energy density $\Phi_s^0(T,D)$ defined in eq.~(\ref{omegas02}) and discussed in \cite{MZ2006}, see also \cite{GD2006}. For $D\gg \lambda$ one finds 
\beq
\label{omtd02}
\Phi_{s}^0(T,D) = 2 \,\sigma_{wg}(T)-\frac{k_{B}T \zeta(3)}{8\pi} \frac{1}{D^2} + o\left(\left(\frac{\lambda}{D}\right)^2\right). 
\eeq
We note the essential difference in the large $D$ behavior of $\Phi_s(T,D)$ and 
$\Phi_{s}^0(T,D)$. In $\Phi_{s}(T,D)$ the leading term 
decays $\sim 1/D$ and is followed by  terms $\sim \ln D/D^2$ and $\sim 1/D^2$, while in $\Phi_{s}^0(T,D)$ the leading term 
decays $\sim 1/D^2$. We also note that the leading decay term in $\Phi_s^0(T,D)/k_{B}T$ has simple form 
$-\zeta(3)/8\pi D^2$ 
often termed in the literature as $-1/D^2$ decay with a universal amplitude $\zeta(3)/8\pi$.  On the other hand, the leading decay term in 
$\Phi_{s}(T,D)/k_{B}T$ is $-\pi \zeta(3/2)/ 4\lambda D$. This $-1/D$  decay cannot be accompanied by a universal amplitude for dimensional reasons. We also note that the term $\sim -1/D^2$ present in $\Phi_{s}(T,D)$ has a different amplitude $\zeta(3)/8\pi - \pi/8 + \pi/8 \ln\left(\pi/4\right)< 0$  from the analogous term in
 $\Phi_s^0(T,D)/k_{B}T$. The difference is, in particular, manifested in their different signs.  
 However, both in $\Phi_{s}(T,D)$ and in $\Phi_{s}^0(T,D)$ the leading terms describing the large distance decay are negative and thus both calculations  agree at least in predicting attractive Casimir forces.
Thus we conclude that in the case of Dirichlet boundary conditions the proper choice of the  path along which the condensed state of ideal Bose gas is approached  reveals a non-universal decay of Casimir forces
\beq
F_{C}(T,D) =  - \frac{\pi k_{B}T}{4 \lambda^3} \left[ \zeta\left(\frac{3}{2}\right) \, \left(\frac{\lambda}{D}\right)^2  \right.
\eeq
\beq
+ \left.2\left(\frac{\lambda}{D}\right)^3 \ln\left(\frac{\lambda}{D}\right) +  
\left(\frac{\zeta(3)}{\pi^2} + \ln\left(\frac{\pi}{4}\right)\right)\,\left(\frac{\lambda}{D}\right)^3 \right]
\eeq
\beq
+ o\left(\left(\frac{\lambda}{D}\right)^3\right).  
\eeq
 With the leading order term $\sim 1/D^2$ followed by $\sim \ln D/D^3$ and $\sim 1/D^3$ it is qualitatively different from the case of periodic or Neumann boundary conditions for which the leading order term is $\sim 1/D^3$.  

\section{Appendix} 
In this Section we sketch the essential steps of calculations leading to our main result in eq.~(\ref{konc01}). 
First, using eq.(\ref{fi02}) we find
\beq
\label{app01}
\frac{\lambda^2}{k_BT}\Phi(T,D)=  \int\limits_{0}^{\infty} dw \sum\limits_{n=1}^{\infty} \ln\left\{1-\exp[-w- s^2(n^2-1)] \right\} = \nonumber
\eeq
\beq 
- \sum\limits_{k=1}^{\infty} \sum\limits_{n=1}^{\infty} \frac{\exp[-k s^2(n^2-1)]}{k^2}
\eeq
The right-hand side of (\ref{app01}) can be rewritten as   $- [S_{1}(s) + S_{2}(s)] $, where
\beq
\label{app03} 
S_{1}(s) = \sum\limits_{k=1}^{\infty} \sum\limits_{n=1}^{\infty} \frac{\exp[-k s^2 n^2]}{k^2} 
\eeq
and
\beq
S_{2}(s) = \sum\limits_{k=1}^{\infty} \frac{\exp(k s^2)-1}{k^2} \sum\limits_{n=1}^{\infty} \exp(-k s^2 n^2).
\eeq 
Using the Poisson formula \cite{MZ2006} we get 
\beq
S_{1}(s) = \sum\limits_{k=1}^{\infty} \frac{1}{k^2} \left[- \frac{1}{2} + \frac{\sqrt{\pi}}{s\sqrt{k}}\left(\frac{1}{2}+\sum\limits_{n=1}^{\infty} \exp[-n^2\pi^2/s^2 k]\right)\right]  \nonumber \\ 
 =\frac{\sqrt{\pi}}{2 } \zeta(5/2) \frac{1}{s}  - \frac{1}{2} \zeta(2) + \frac{\zeta(3)}{2 \pi^2}{s^2} + \dots \,.
\eeq
The term $S_{2}(s)$ can be rewritten as
\beq
\label{app10}
S_{2}(s)= \sum\limits_{k=1}^{\infty} \left[\frac{s^2}{k} + \frac{s^4}{2!} + \frac{s^6}{3!} k + \dots \right] \sum\limits_{n=1}^{\infty} 
\exp(-k s^2 n^2) = \nonumber 
\eeq 
\beq  
-s^2 \sum\limits_{n=1}^{\infty} \ln(1- \exp(-s^2 n^2) + \frac{s^4}{2!} \sum\limits_{n=1}^{\infty} \frac{1}{\exp(s^2n^2)-1} + \nonumber 
\eeq
\beq
\frac{s^6}{3!} \sum\limits_{n=1}^{\infty} \frac{\exp(s^2n^2)}{[\exp(s^2n^2)-1]^2} + \dots
\eeq
In order to evaluate the first term on the rhs of eq.~(\ref{app10}) one uses the Euler-Maclaurin formula \cite{ZUK1977,NP2014}
\beq
\label{MacL01}
\sum\limits_{n=1}^{N} f(n) = \int\limits_{0}^{N} dx f(x) + \frac{1}{2} \left[ f(N) -f(0)\right]  +  \nonumber
\eeq
\beq 
2\,\sum\limits_{p=1}^{N}\int\limits_{0}^{N} dx f(x) \cos(2\pi p x)
\eeq
which, after inserting $f(x) = \ln\left[1- \exp(-s^2 x^2)\right] - \ln x^2 $, gives 
\beq
\label{app11}
\sum\limits_{n=1}^{\infty} \ln[1- \exp(-s^2 n^2)] = \nonumber 
\eeq
\beq
- \ln s -\frac{\sqrt{\pi}}{2 s} \zeta(3/2) + \ln(2\pi) +O(e^{-c/s}) ,
\eeq
where $c>0$. The remaining terms on the rhs of eq.~(\ref{app10}) are at least of order $s^2$. 
After summing all contributions of order $s^2$ one obtains 
\beq
 \frac{s^4}{2!} \sum\limits_{n=1}^{\infty} \frac{1}{\exp(s^2n^2)-1} + 
\frac{s^6}{3!} \sum\limits_{n=1}^{\infty} \frac{\exp(s^2n^2)}{[\exp(s^2n^2)-1]^2} + \dots \nonumber \\ 
 = s^2 \, \sum\limits_{n=1}^{\infty} \frac{\zeta(2n)}{n(n+1)} + o(s^2) = s^2 \ln (2\pi) - \frac{s^2}{2} + o(s^2). 
\eeq
The above expressions for $S_1(s)$ and $S_2(s)$ give the final result in eq.~(\ref{konc01}).

\end{document}